\documentclass[letterpaper,1p]{elsarticle}

\graphicspath{{.}}
\DeclareGraphicsExtensions{.pdf,.jpeg,.jpg,.png}

\usepackage[cmex10]{amsmath}
\usepackage{url}

\usepackage{multirow}
\usepackage{arydshln} 

\usepackage{subfig}

\begin{document}

\begin{frontmatter}
\title{Towards Quality of Experience Determination for  Video in Augmented Binocular Vision Scenarios}

\author[rvt]{Patrick~Seeling}
\ead{patrick.seeling@cmich.edu}
\ead[url]{http://patrick.seeling.org}
\address[rvt]{Department of Computer Science, Central Michigan University, Mount Pleasant, MI 48859, USA.}

\begin{abstract}
	\boldmath
With the continuous growth in the consumer markets of mobile smartphones and increasingly in augmented binocular vision wearable devices, several avenues of research investigate the relationships between the quality perceived by mobile users and the delivery mechanisms at play to support a high quality of experience for mobile users.
In this paper, we present the first study that evaluates the relationships of mobile movie quality and the viewer--perceived quality thereof in an augmented binocular vision setting employing commercially available head--mounted see--through devices.
We find that participants tend to overestimate the video quality when compared to a scaled representation and exhibit a significant variation of accuracy that leans onto the movie content and its dynamics.
Our findings, thus, can broadly impact future media adaptation and delivery mechanisms for this new display format of mobile multimedia and spur follow--up research in this increasingly popular domain.
\end{abstract}


\end{frontmatter}

\section{Introduction}

In recent years, the amount of connected devices that are carried by mobile users has increased drastically and will become one of the dominant drivers for future mobile networking, as described by Cisco, Inc.~\cite{CiscoInc:2014td}.
A secondary forecasted trend is the continuously large fraction of mobile data that is required due to multimedia consumption while users are ``on--the--go.''
While currently, smartphones and tablet computers are the dominant form of media consumption and display, the prospect of reality--augmenting wearable devices will likely account for a significant portion of the interaction with mobile multimedia content in future immersive communications systems~\cite{Apostolopoulos:2012ep}.
Augmented reality has been an area of research in ubiquitous computing for some time~\cite{Mann:1994ur} and is subject to ongoing research efforts~\cite{Zhou:2008to}.
Several application scenarios were evaluated in recent years in different domains, such as smartphone--based information overlay systems~\cite{Takacs:2008wo,Wither:2011gx}, outdoor systems with multiple elements~\cite{Gleue:2001wn,Ribo:2002th}, navigation~\cite{Castle:2008vs}, or general information systems combining both~\cite{Sulisz:2012da}.

Several industry-based solutions were developed recently in parallel to augmented reality devices, which target the multimedia playback application scenario in the predominantly consumer market space.
Augmented reality devices that are performing in a heads-up-display (HUD) or Head--Mounted Display (HMD) manner are increasingly targeting the professional and consumer application space alike, indicating future broad adaptation.
While these types of devices are available in a broad variety of implementations (see, e.g., \cite{VanKrevelen:2010vm} for an overview of different types), a slow convergence of systems has begun, especially in the consumer space.
Examples for current commercial off-the--shelf (COTS) devices available include the Occulus Rift, Sony HMZ-T1 Personal 3D Viewer, Epson Moverio BT-100, or Google Glasses.
We note that only the latter two are optical see--through devices and thus similar to the one presented in~\cite{Kanbara:2000wu}, showcasing how these device types have undergone additional improvements and are now consumer--graded.

The evaluation of these types of systems and related issues have attracted different research groups and a recent survey~\cite{Kruijff:2010vy} highlights ongoing issues for the various system types. 
Evaluations performed also target the user--perception of augmentation for daily life scenarios, such as in~\cite{Bonanni:2005wt}, or how to limit the amount of additional information, as in, e.g., \cite{Kalkofen:2007vh}.
Perceptual evaluations oftentimes consider the segmentation of virtualized/augmented items, such as in, e.g., \cite{Sanches:2012dc}.
There are several constraints that have to be considered in this particular scenario, especially from a communications point of view, when targeting the delivery of video data to these types of systems, as the replication of video content with natural features might behave significantly different from overlaid computer--generated information. 
In~\cite{Perritaz:2007dn}, the authors evaluate an industrial system that consists of an opaque HMD that displays video sequences at different target bit rates (and resulting imperfection or distortion levels) and user select different encoder properties at the target rate--points, resulting in a combination of frame rate and compression. 
Our evaluation is significantly different in that we provide participating users with a see--trough COTS HMD at prescribed video frame rates.
Significant differences can be expected for the perceived video quality based on the type of the visual display~\cite{Kruijff:2010vy}.

In this paper, we investigate the applicability of existing video quality metrics, such as the frequently used Peak Signal--to--Noise Ratio (PSNR), Structural Similarity Index Metric (SSIM), and Video Quality Metric (VQM), in the augmented reality space and correlate encoded video quality with subjectively rated perceived video quality levels.
The perceptual video quality is measured using mean opinion scores (MOS) obtained from multiple human test subjects according to~\cite{ITUR:2012vm}.
We note that currently, no specific testing standard has been established for the determination or evaluation of audio--visual objective or perceptual qualities employing COTS see--through devices or for augmented reality settings.
In turn, we consider the existing standard as outlined in recommendation ITU--T BT.500--13 \cite{ITUR:2012vm} as a general guideline for the experiments we perform here.
In this seminal work, we provide a high--level overview of subjective quality ratings for longer movie segments performing an initial view at the underlying characteristics at play.

The broad potential for implementation in future systems that contain augmented single view or binocular vision display modalities is manifold, as media adaptations for specific video material might benefit content and network providers while maintaining a sufficiently high level of quality of experience~\cite{Nam:2005uy}.
Here, we focus on the evaluation of video compression quality without potentially lossy network transport when viewed in a scenario where binocular vision is augmented.

In the remainder of this contribution, we initially describe the measurement setup (including the wearable device, the developed mobile application, and the encoded video sequence characteristics) in greater detail.
We continue with a detailed description of the obtained results in Section~\ref{s:results} and evaluate the participating users' video quality selection performance in Section~\ref{s:selection}. 
We conclude with an outlook on future activities in Section~\ref{s:conc}.

\section{Measurement Setup}
\label{s:setp}
In this section, we initially describe the employed wearable head--mounted optical see--through display and the application we developed for the experimentation. 
We continue with a description of media characteristics and performed experimentation with volunteering participants. 

\subsection{Binocular Augmented Vision Heads--Up Display}
\begin{figure}
\centering
\includegraphics[width=0.67\linewidth]{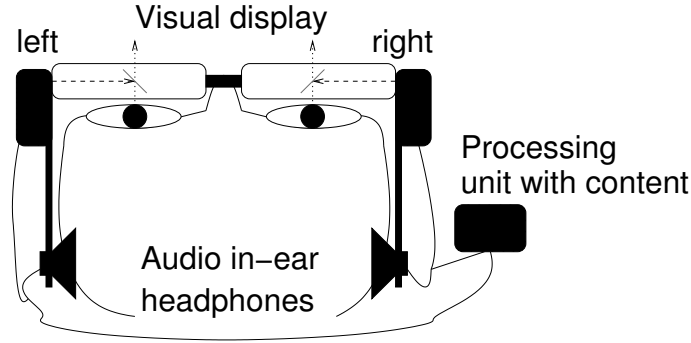}
\caption{Schematic view of the head--wearable binocular see--through glassses with in--ear headphones and the main processing unit that additionally stores the audio--visual content for local playout.}
\label{fig:viewer}
\end{figure}

We employed the Epson Moverio BT--100 mobile viewer, which consists of a wearable 3D--capable heads--up display unit and a central processing unit. 
The processing unit features both a directional pad and a touch pad and employs the Android Operating System version 2.2 (``Froyo''). 
The unit is connected via wires to deliver video signals and power to the see--through display unit, with a display control being integrated into the wired connection. 
We illustrate the overall system configuration in Figure~\ref{fig:viewer}, noting that no networking is involved for the display of the content, as it is contained on the processing unit.

The display unit has a resolution of  $960 \times 540$ pixels of 24--bit color with LED light sources and a 23 degree field of view. 
Without the additionally available shade, a maximum of 70 \% transparency is realized for the display.
The images are projected from a display panel built into the device's sides, from which light is reflected through a lens, and in turn hits a half--mirror layer in the light guide material.
As we consider an evaluation of a commercially available off-the--shelf augmented binocular vision HMD, no specific calibration was performed.
Factory settings were applied, which allow for 24-bit color reproduction at 60 Hz and the built--in LED light intensity was set to maximum for highest contrast.

\subsection{Mobile Application}
\begin{figure}
\centering
\includegraphics[width=0.95\linewidth]{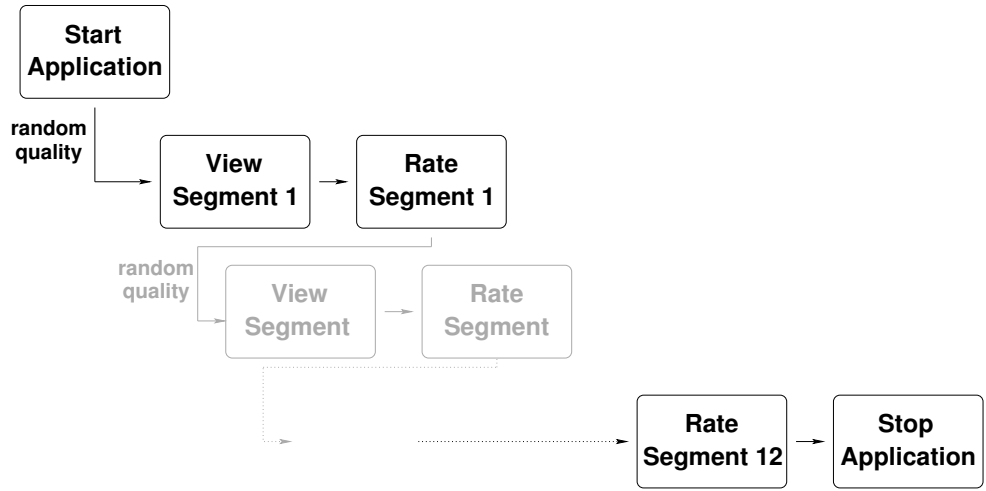}
\caption{Overall flow of the mobile application: Participants are presented with a random selected quality for each segment and asked for a rating at the end of the segment until all segments are played out.}
\label{fig:app}
\end{figure}

We developed a mobile Android application that can be executed on the wearable display's control unit.
The application displays a movie sequence (video and audio content), followed by a Likert--scale question to rate the quality of the last displayed sequence.
We illustrate this approach in Figure~\ref{fig:app}.

Initially, a random quality for a movie sequence is selected and its value stored in a text file on--device before starting the audio/video play--out from the on--device storage, disabling potential network transmission impacts.
After play--out, the user is asked to select a quality level from a presented Likert--scale, with each selection of a quality level captured in the same text file on--device.
We do not enforce a time limit for the rating procedure, as users need to interface with the mobile application through the processing unit's touchpad.
	Adding a time limit at this stage would increase the potential stress on participants as they make their selections, in turn potentially influencing their ratings.
This process continues until all movie sequences are played out.
The created text file with the randomly chosen movie qualities and user rankings of movie qualities can afterwards be copied from the device to a desktop computer for further processing.

\subsection{Multimedia Description}
\label{ss:vieo_desc}
We employ the publicly available and Creative Commons licensed \emph{Tears of Steel} short movie as source, which depicts an epic struggle between humans and robots in the future.
The video was made by the Blender foundation, merging computer--generated graphics generated by the the open--sourced Blender software with real--world filmed scenes in Amsterdam, The Netherlands.
We employ this short film as representative of today's video contents which commonly feature a combination of real--world and computer--generated source materials  (we refer the interested reader to \url{http://tearsofsteel.org} for more details about the movie).

\begin{figure}
\centering
\includegraphics[width=0.75\linewidth]{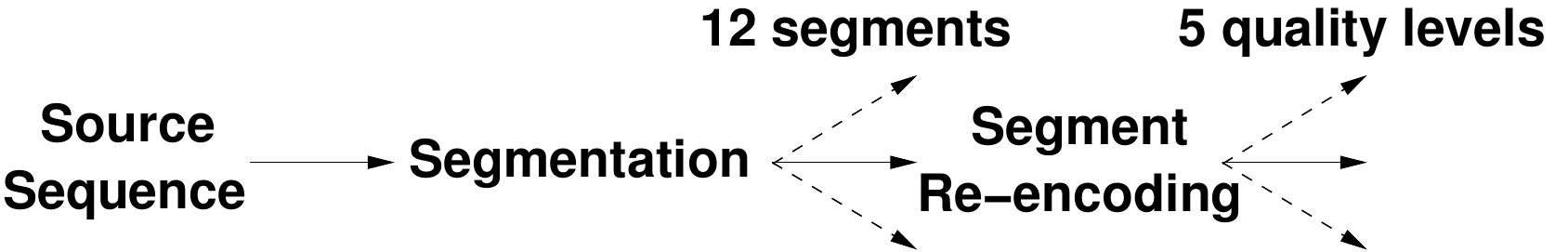}
\caption{Segmentation and encoding of the original video sequence resulting in 12 segments, each encoded into five different quality levels.}
\label{fig:segmentation}
\end{figure}

We employ the publicly available 720p version of the movie and segment this source into logically connected scenes for processing, as illustrated in Figure~\ref{fig:segmentation}.
The individual video segments were resized to support the native $960 \times 540$ resolution of the augmented binocular vision glasses and re--encoded using the popular open--source \texttt{ffmpeg} video tool in 24 frames per second.
The encoder used was \texttt{libx264} with constant rate factor (also known as constant quantization scale factor) settings of 1, 30, 35, 40, and 45, resulting in a constant quality encoding with variable bitrates.
We selected this approach, as the source video sequence itself was professionally encoded with high fidelity settings, and minimal encoding losses; these are contained in our encodings as well.
The output was visually inspected to ensure that the settings provided significant differences in visual quality to allow mapping to a quality scale from 1--5, respectively.
This represents quality level differences observable within typical streaming scenarios; the resulting values for the PSNR as an objective video quality metric comparing the source video quality to the encoded video are provided in Table~\ref{tab:psnr}.
\begin{table*}
\centering
\caption{Overview of characteristics for the \emph{Tears of Steel} movie used for experimentation segmented into shorter segments at different quality levels. The typical difference between quality levels is 3 dB, starting with visually identifiable encoding losses in the second quality level.}

\tiny
\begin{tabular}{|c|c|c||c|c|c|c|c|}\hline

Segment & Frames & Level & \multicolumn{5}{|c|}{PSNR}\\
				&          &         & $q^{\min}_s$ [dB] & $\overline{q}_s$ [dB]& $q^{\max}_s$ [dB]& $\sigma(q_s)$ [dB]& $\mathrm{CoV}(q_s)$ \\\hline \hline

\multirow{5}{*}{1}  & \multirow{5}{*}{1548} & 1 & 52.497 & 64.350 & 188.131 & 27.402 & 0.426\\\cdashline{3-8}[1pt/1pt]
 &  & 2 & 34.846 & 45.035 & 188.131 & 25.105 & 0.557\\\cdashline{3-8}[1pt/1pt] 
 &  & 3 & 32.056 & 42.205 & 188.131 & 25.619 & 0.607\\\cdashline{3-8}[1pt/1pt] 
 &  & 4 & 29.391 & 39.375 & 188.131 & 26.101 & 0.663\\\cdashline{3-8}[1pt/1pt] 
 &  & 5 & 26.715 & 36.380 & 188.131 & 26.405 & 0.726\\\hline \hline

\multirow{5}{*}{2}  & \multirow{5}{*}{1327} & 1 & 53.490 & 58.963 & 188.131 & 5.713 & 0.097\\\cdashline{3-8}[1pt/1pt] 
 & & 2 & 35.835 & 41.521 & 188.131 & 5.174 & 0.125\\\cdashline{3-8}[1pt/1pt] 
 & & 3 & 33.090 & 38.537 & 188.131 & 5.191 & 0.135\\\cdashline{3-8}[1pt/1pt] 
 & & 4 & 30.292 & 35.573 & 188.131 & 5.226 & 0.147\\\cdashline{3-8}[1pt/1pt] 
 & & 5 & 27.247 & 32.500 & 62.358  & 3.096 & 0.095\\\hline \hline

\multirow{5}{*}{3}  & \multirow{5}{*}{1259} & 1 & 53.333 & 57.216 & 65.594 & 2.753 & 0.048\\\cdashline{3-8}[1pt/1pt] 
 & & 2 & 36.237 & 40.371 & 44.755 & 1.444 & 0.036\\\cdashline{3-8}[1pt/1pt] 
 & & 3 & 33.307 & 37.504 & 41.913 & 1.539 & 0.041\\\cdashline{3-8}[1pt/1pt] 
 & & 4 & 30.428 & 34.572 & 38.952 & 1.612 & 0.047\\\cdashline{3-8}[1pt/1pt] 
 & & 5 & 27.736 & 31.531 & 35.939 & 1.627 & 0.052\\\hline \hline

\multirow{5}{*}{4}  & \multirow{5}{*}{823} & 1 & 52.859 & 56.537 & 65.128 & 2.926 & 0.052\\\cdashline{3-8}[1pt/1pt] 
 & & 2 & 34.397 & 39.234 & 43.729 & 1.742 & 0.044\\\cdashline{3-8}[1pt/1pt] 
 & & 3 & 31.589 & 36.347 & 41.049 & 1.853 & 0.051\\\cdashline{3-8}[1pt/1pt] 
 & & 4 & 29.030 & 33.426 & 38.176 & 1.895 & 0.057\\\cdashline{3-8}[1pt/1pt] 
 & & 5 & 26.379 & 30.454 & 35.215 & 1.809 & 0.059\\\hline \hline

\multirow{5}{*}{5}  & \multirow{5}{*}{1227} & 1 & 52.084 & 56.091 & 65.310 & 2.856 & 0.051\\\cdashline{3-8}[1pt/1pt] 
 & & 2 & 33.364 & 37.876 & 44.670 & 1.465 & 0.039\\\cdashline{3-8}[1pt/1pt] 
 & & 3 & 30.617 & 34.911 & 41.981 & 1.514 & 0.043\\\cdashline{3-8}[1pt/1pt] 
 & & 4 & 27.685 & 32.015 & 38.688 & 1.518 & 0.047\\\cdashline{3-8}[1pt/1pt] 
 & & 5 & 24.963 & 29.164 & 34.953 & 1.460 & 0.050\\\hline \hline

\multirow{5}{*}{6}  & \multirow{5}{*}{1699} & 1 & 51.612 & 55.879 & 65.490 & 2.733 & 0.049\\\cdashline{3-8}[1pt/1pt] 
 & & 2 & 33.492 & 38.527 & 44.238 & 2.179 & 0.057\\\cdashline{3-8}[1pt/1pt] 
 & & 3 & 30.947 & 35.641 & 41.965 & 2.455 & 0.069\\\cdashline{3-8}[1pt/1pt] 
 & & 4 & 28.399 & 32.735 & 39.661 & 2.638 & 0.081\\\cdashline{3-8}[1pt/1pt] 
 & & 5 & 26.046 & 29.793 & 37.143 & 2.696 & 0.091\\\hline \hline

\multirow{5}{*}{7}  & \multirow{5}{*}{1308} & 1 & 52.164 & 56.742 & 65.984 & 2.788 & 0.049\\\cdashline{3-8}[1pt/1pt] 
 & & 2 & 36.311 & 39.802 & 44.135 & 1.551 & 0.039\\\cdashline{3-8}[1pt/1pt] 
 & & 3 & 33.425 & 36.977 & 41.526 & 1.717 & 0.046\\\cdashline{3-8}[1pt/1pt] 
 & & 4 & 30.322 & 34.084 & 38.752 & 1.821 & 0.053\\\cdashline{3-8}[1pt/1pt] 
 & & 5 & 27.385 & 31.126 & 35.924 & 1.838 & 0.059\\\hline \hline

\multirow{5}{*}{8}  & \multirow{5}{*}{1160} & 1 & 52.162 & 56.562 & 65.327 & 2.812 & 0.050\\\cdashline{3-8}[1pt/1pt] 
 & & 2 & 33.748 & 40.078 & 43.457 & 1.493 & 0.037\\\cdashline{3-8}[1pt/1pt] 
 & & 3 & 30.807 & 37.318 & 40.735 & 1.608 & 0.043\\\cdashline{3-8}[1pt/1pt] 
 & & 4 & 27.779 & 34.425 & 37.905 & 1.675 & 0.049\\\cdashline{3-8}[1pt/1pt] 
 & & 5 & 25.204 & 31.338 & 34.926 & 1.615 & 0.052\\\hline \hline

\multirow{5}{*}{9}  & \multirow{5}{*}{1737} & 1 & 52.168 & 56.634 & 66.097 & 2.235 & 0.039\\\cdashline{3-8}[1pt/1pt] 
 & & 2 & 32.315 & 38.736 & 47.341 & 2.230 & 0.058\\\cdashline{3-8}[1pt/1pt] 
 & & 3 & 29.132 & 35.723 & 45.146 & 2.287 & 0.064\\\cdashline{3-8}[1pt/1pt] 
 & & 4 & 26.166 & 32.573 & 42.381 & 2.210 & 0.068\\\cdashline{3-8}[1pt/1pt] 
 & & 5 & 23.447 & 29.405 & 38.572 & 2.023 & 0.069\\\hline \hline

\multirow{5}{*}{10}  & \multirow{5}{*}{817} & 1 & 51.994 & 56.337 & 65.958 & 3.067 & 0.054\\\cdashline{3-8}[1pt/1pt] 
 & & 2 & 34.013 & 38.588 & 43.726 & 2.076 & 0.054\\\cdashline{3-8}[1pt/1pt] 
 & & 3 & 30.999 & 35.700 & 40.875 & 2.283 & 0.064\\\cdashline{3-8}[1pt/1pt] 
 & & 4 & 28.087 & 32.801 & 37.993 & 2.402 & 0.073\\\cdashline{3-8}[1pt/1pt] 
 & & 5 & 25.029 & 29.916 & 35.196 & 2.424 & 0.081\\\hline \hline

\multirow{5}{*}{11}  & \multirow{5}{*}{1216} & 1 & 52.913 & 57.439 & 188.131 & 11.682 & 0.203\\\cdashline{3-8}[1pt/1pt] 
 & & 2 & 35.194 & 39.431 & 188.131 & 12.984 & 0.329\\\cdashline{3-8}[1pt/1pt] 
 & & 3 & 32.101 & 36.529 & 188.131 & 13.249 & 0.363\\\cdashline{3-8}[1pt/1pt] 
 & & 4 & 29.068 & 33.671 & 188.131 & 13.503 & 0.401\\\cdashline{3-8}[1pt/1pt] 
 & & 5 & 26.274 & 30.813 & 188.131 & 13.751 & 0.446\\\hline \hline

\multirow{5}{*}{12}  & \multirow{5}{*}{3499} & 1 & 49.845 & 61.468 & 188.131 & 25.622 & 0.417\\\cdashline{3-8}[1pt/1pt] 
 & & 2 & 30.756 & 41.336 & 188.131 & 29.331 & 0.710\\\cdashline{3-8}[1pt/1pt] 
 & & 3 & 26.764 & 38.066 & 188.131 & 30.013 & 0.788\\\cdashline{3-8}[1pt/1pt] 
 & & 4 & 23.177 & 35.026 & 188.131 & 30.648 & 0.875\\\cdashline{3-8}[1pt/1pt] 
 & & 5 & 19.929 & 32.158 & 188.131 & 31.223 & 0.971\\\hline 
\end{tabular}
\label{tab:psnr}
\end{table*}

The average segment length is 1486 frames, with the shortest segment covering 817 frames and the longest segment covering 3499 frames.
(We note that the individual lengths are longer than the common 10 s employed in current perceptual video quality evaluations, as the goal of this evaluation is a high--level content overview, leaving a more detailed evaluation for future works.)
Segments 4 and 10 contain significantly less frames than the other segments; this was required to group multiple scenes into logically fitting segment enabling end of segment questioning about quality.
The longest segment of the movie is the last one, which includes the titles and a short end sequence.
The average quality in each segment $s$, denoted as $\overline{q}_s$  and measured as averaged PSNR of the video frames within the sequence, is above 55 dB for the highest quality level and just above 29 dB for the lowest.
The largest difference typically is encountered from the highest quality level to the second one, representing the introduction of visually recognizable encoding losses.
Afterwards, the difference between the different quality levels is around 3 dB.
Comparing the variability of the individual video frame $i$ quality $q_{i,s}$ in the different segments, either as standard deviation $\sigma(q_s)$ or coefficient of variation $\mathrm{CoV}(q_s)$, we observe a significantly higher level for the first and last two segments of the video.
The reason for this increased level is the number of all--black and title/credit content video frames that are encountered in the beginning of the movie and towards the end.
The homogeneous single--color content increases the coding efficiency and results in no measurable coding losses, bringing the PSNR to an increased level (indicated by the maximum video frame quality value $q_s^{\max}$).

\begin{figure}[]
	\centering
    \subfloat[SSIM\label{sfig:ssim}]{%
      \includegraphics[width=0.75\textwidth]{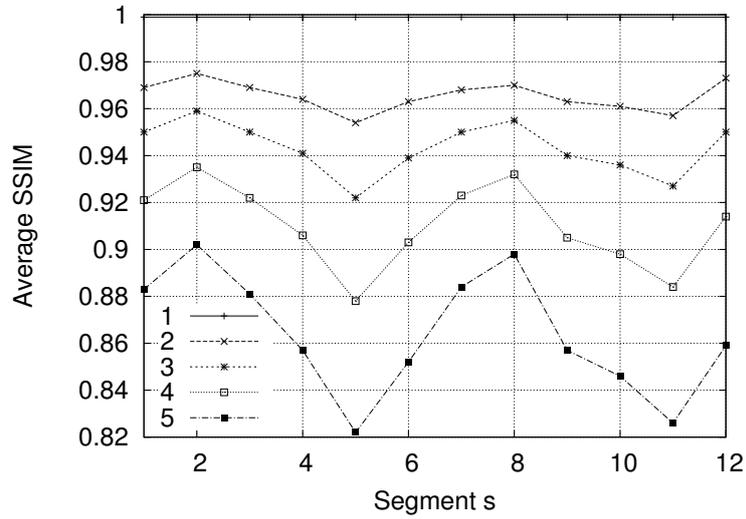}
    }
    \\
    \subfloat[VQM\label{sfig:vqm}]{%
      \includegraphics[width=0.75\textwidth]{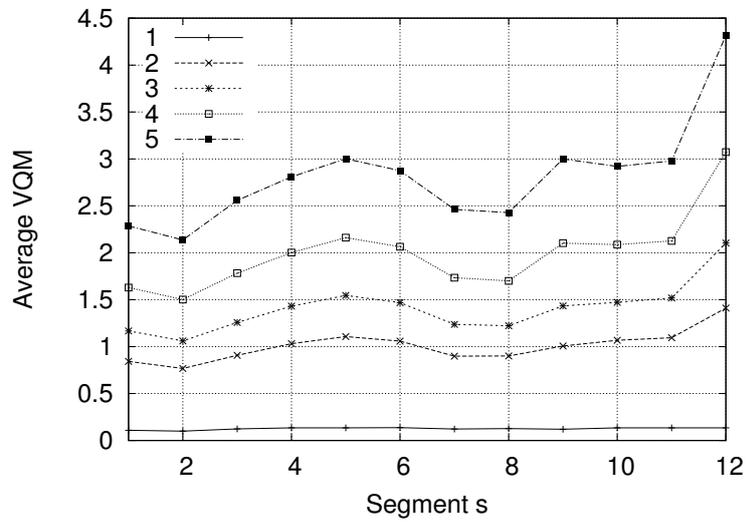}
    }
    \caption{Segment averages of the Structured Similarity Index Metric (SSIM) and Video Quality Metric (VQM).}
	\label{fig:ssimvqm}
\end{figure}

As the PSNR is not the most correlated to user--perceived qualities, albeit a frequently used one, especially in network performance and video coder evaluations, we additionally provide the Structured Similarity Index Metric (SSIM) and Video Quality Metric (VQM) as additional objective video quality evaluation metrics, see, e.g., \cite{SSIM,VQM}.
We illustrate the segment averages of these two metrics in Figure~\ref{fig:ssimvqm}.

Focusing on the average SSIM initially, we note that the highest values represent the smallest encoding losses.
The higher encoding quantization scales (or constant rate factors) result in varying SSIM averages, based on the individual segment content.
Segments 5 and 11 exhibit the overall lowest average SSIM values obtained for each encoding setting within our setup.
In addition, we observe that the changes between the individual segments' different encodings are relatively comparable.
For the segment average VQM values in Figure~\ref{sfig:vqm}, we observe a similar, albeit reverse trend due to the VQM measuring the visual impairments.
We again note that scenes 5 and 11 exhibit higher values, but the VQM average for the final scene 12 exhibits the highest level of approximated visual impairments.
We again note that the lowest encoder quantization scale results in only negligible quality impairments as identified by the VQM.
Corroborating the observation for the SSIM, we find that the relative changes in between the different settings for the individual segments are very comparable.

As the perceived multimedia quality is the result of audio--visual stimuli interplays~\cite{Tasaka:2006}, 
we note that we do not process the audio component of the movie segments and copy the original audio source to the various new segment versions.
The original source audio is 48 kHz sampled stereo and encoded in the MPEG Advanced Audio Coding (AAC) standard.
Though this encoding standard is lossy, the 128 kbps used for compression of the audio are common in consumer applications and represent a common quality level that should not impact the overall perception.

\subsection{Experimental Set--Up}
Original research protocol submission to the Institutional Review Board (IRB) at Central Michigan University was performed in February, 2014 and approval was obtained beginning of April, 2014.
Participating volunteers were recruited from students, faculty, and staff of the Department of Computer Science at Central Michigan University from April through May, 2014.
The participants were instructed about the nature of the experiment and its overall procedural flow; this was followed by a description of the wearable display and the required interaction with it.
The instruction part was followed by consent form administration before fitting the wearable display and commencing experimentation.
The experiments took place in well--lit office spaces and classrooms, with participants being instructed to look at a stretch of white wall or a white--board to allow for comparability of results.
All of the participants used the in--ear headphones to play back audio accompanying the visual content to be evaluated.

\section{Experimental Results}
\label{s:results}
In this section, we discuss the results obtained through the experimentation with participants. 
We report findings for experiments conducted with 15 volunteers, which is the required sample size for audio--visual experiments outlined in~\cite{ITUR:2012vm}. 
We note, however, that the experimental design resulted in not every segment encoding level having the same number of viewers.
We present the encoded video quality levels $v$, which have been ranked based on the average segment PSNR as one potential approach, as in Table~\ref{tab:psnr}. We subsequently present the participant--selected qualities $p$ for each user $u$ and segment $s$, both are provided in Table~\ref{tab:selections}.

\begin{table}[]
	\centering
	\caption{Experimental results for encoded video quality levels $v$ and the participant--selected qualities $p$ for each user $u$ and segment $s$.}
	\begin{tabular}{|ll||c|c|c|c|c|c|c|c|c|c|c|c|c|c|c|}\hline
		Segm. 			& Mode		& \multicolumn{15}{|c|}{User $u$} \\
		$s$ 			  & $v/p$ 	& 1     & 2     & 3     & 4     & 5     & 6     & 7     & 8     & 9     & 10    & 11    & 12    & 13    & 14    & 15 \\\hline\hline
		
		\multirow{2}{*}{1} & $v$ & 2     & 1     & 1     & 3     & 5     & 1     & 5     & 3     & 5     & 1     & 3     & 5     & 5     & 5     & 1 \\\cdashline{2-17}[1pt/1pt] 
		&$p$& 3     & 3     & 1     & 2     & 4     & 2     & 4     & 4     & 3     & 3     & 5     & 3     & 3     & 5     & 2 \\\hline
		
		\multirow{2}{*}{2}  &$v$& 2     & 5     & 4     & 5     & 4     & 3     & 5     & 4     & 1     & 1     & 5     & 1     & 4     & 1     & 1 \\\cdashline{2-17}[1pt/1pt] 
		&$p$& 4     & 4     & 4     & 5     & 4     & 2     & 2     & 3     & 2     & 3     & 4     & 1     & 4     & 4     & 3 \\\hline
		
		\multirow{2}{*}{3}  &$v$& 4     & 2     & 1     & 5     & 1     & 5     & 3     & 3     & 3     & 3     & 1     & 2     & 1     & 4     & 3 \\\cdashline{2-17}[1pt/1pt] 
		&$p$& 5     & 3     & 1     & 5     & 2     & 3     & 2     & 4     & 3     & 4     & 1     & 2     & 1     & 5     & 4 \\\hline
		
		\multirow{2}{*}{4}  &$v$& 3     & 1     & 4     & 1     & 5     & 4     & 3     & 4     & 5     & 3     & 3     & 5     & 1     & 5     & 5 \\\cdashline{2-17}[1pt/1pt] 
		&$p$& 5     & 2     & 4     &       & 5     & 3     & 3     & 4     & 4     & 4     & 2     & 3     & 1     & 5     & 5 \\\hline
		
		\multirow{2}{*}{5}  &$v$& 5     & 1     & 1     & 2     & 1     & 1     & 2     & 3     & 4     & 3     & 4     & 3     & 3     & 3     & 3 \\\cdashline{2-17}[1pt/1pt] 
		&$p$& 5     & 3     & 2     & 3     & 1     & 2     & 1     & 3     & 3     & 4     & 2     & 3     & 2     & 5     & 4 \\\hline
		
		\multirow{2}{*}{6}  &$v$& 3     & 5     & 2     & 2     & 2     & 3     & 4     & 3     & 4     & 4     & 1     & 3     & 3     & 3     & 5 \\\cdashline{2-17}[1pt/1pt] 
		&$p$& 5     & 3     & 3     & 3     & 3     & 2     & 2     & 4     & 4     & 4     & 2     & 3     & 1     & 5     & 4 \\\hline
		
		\multirow{2}{*}{7}  &$v$& 1     & 2     & 5     & 1     & 1     & 2     & 2     & 3     & 1     & 5     & 4     & 2     & 1     & 5     & 4 \\\cdashline{2-17}[1pt/1pt] 
		&$p$& 2     & 3     & 4     & 3     & 2     & 2     & 2     & 5     & 1     & 4     & 3     & 2     & 1     & 5     & 4 \\\hline
		
		\multirow{2}{*}{8}  &$v$& 3     & 5     & 4     & 2     & 2     & 1     & 1     & 2     & 3     & 1     & 4     & 2     & 1     & 2     & 2 \\\cdashline{2-17}[1pt/1pt] 
		&$p$& 5     & 4     & 3     & 3     & 3     & 1     & 1     & 2     & 3     & 2     & 3     & 2     & 1     & 5     & 3 \\\hline
		
		\multirow{2}{*}{9}  &$v$& 3     & 3     & 3     & 1     & 3     & 4     & 4     & 1     & 1     & 5     & 1     & 4     & 2     & 2     & 2 \\\cdashline{2-17}[1pt/1pt] 
		&$p$& 5     & 3     & 3     & 1     & 3     & 4     & 4     & 1     & 1     & 5     & 3     & 4     & 1     & 5     & 3 \\\hline
		
		\multirow{2}{*}{10}  &$v$& 4     & 1     & 3     & 1     & 3     & 2     & 5     & 1     & 1     & 5     & 5     & 5     & 5     & 5     & 3 \\\cdashline{2-17}[1pt/1pt] 
		&$p$& 5     & 3     & 3     & 1     & 2     & 3     & 4     & 1     & 2     & 4     & 3     & 3     & 4     & 5     & 4 \\\hline
		
		\multirow{2}{*}{11} &$v$& 1     & 3     & 4     & 4     & 4     & 5     & 4     & 4     & 5     & 4     & 4     & 2     & 5     & 2     & 1 \\\cdashline{2-17}[1pt/1pt] 
		&$p$& 2     & 3     & 5     & 5     & 3     & 4     & 4     & 4     & 5     & 4     & 3     & 2     & 4     & 5     & 3 \\\hline
		
		\multirow{2}{*}{12} &$v$& 3     & 2     & 4     & 2     & 4     & 3     & 1     & 1     & 4     & 3     & 4     & 1     & 3     & 2     & 4 \\\cdashline{2-17}[1pt/1pt] 
		&$p$& 3     &       & 5     & 4     & 4     & 3     & 3     & 2     & 3     & 3     & 1     & 1     & 2     & 4     & 5 \\\hline
		
	\end{tabular}%
	\label{tab:selections}%
\end{table}%
The overall average for the randomly chosen video quality levels is $\mu_v=2.94$ with a standard deviation of $\sigma_v=1.44$, while the user--selected ones exhibit a higher average with slightly smaller variability ($\mu_p=3.13, \sigma_p=1.24$).
	The differences between the two indicates that users overall select slightly higher quality levels than actually encoded, based on our initial ranking scale.

\subsection{Mean Opinion Scores}
\begin{figure}
	\centering
	\includegraphics[width=0.85\linewidth]{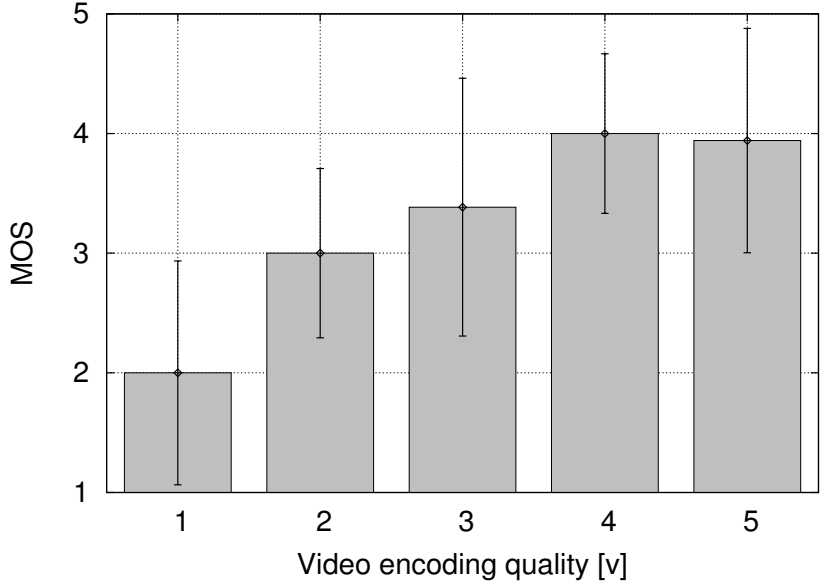}
	\caption{Mean Opinion Score (MOS) ratings related to identified video quality levels.}
	\label{fig:mos_overall}
\end{figure}

We illustrate the mean opinion scores from all video segments and users based on our video encoding ranking in Figure~\ref{fig:mos_overall}.

We observe that our initial quality ranking from 1--5 is reflected trend--wise in the user selection. 
We note, however, that the mean user selection score begins at an average of 2, which is higher than our initial ranking of video qualities.
The increased video qualities result in a progressive, almost logarithmic increase as our ranking increases.
We also note that the individual ratings at each of the quality rankings exhibit fairly high levels of variability in between participants.

\begin{figure}[]
	\centering
	\subfloat[PSNR\label{sfig:mospsnr}]{%
		\includegraphics[height=0.3\textheight]{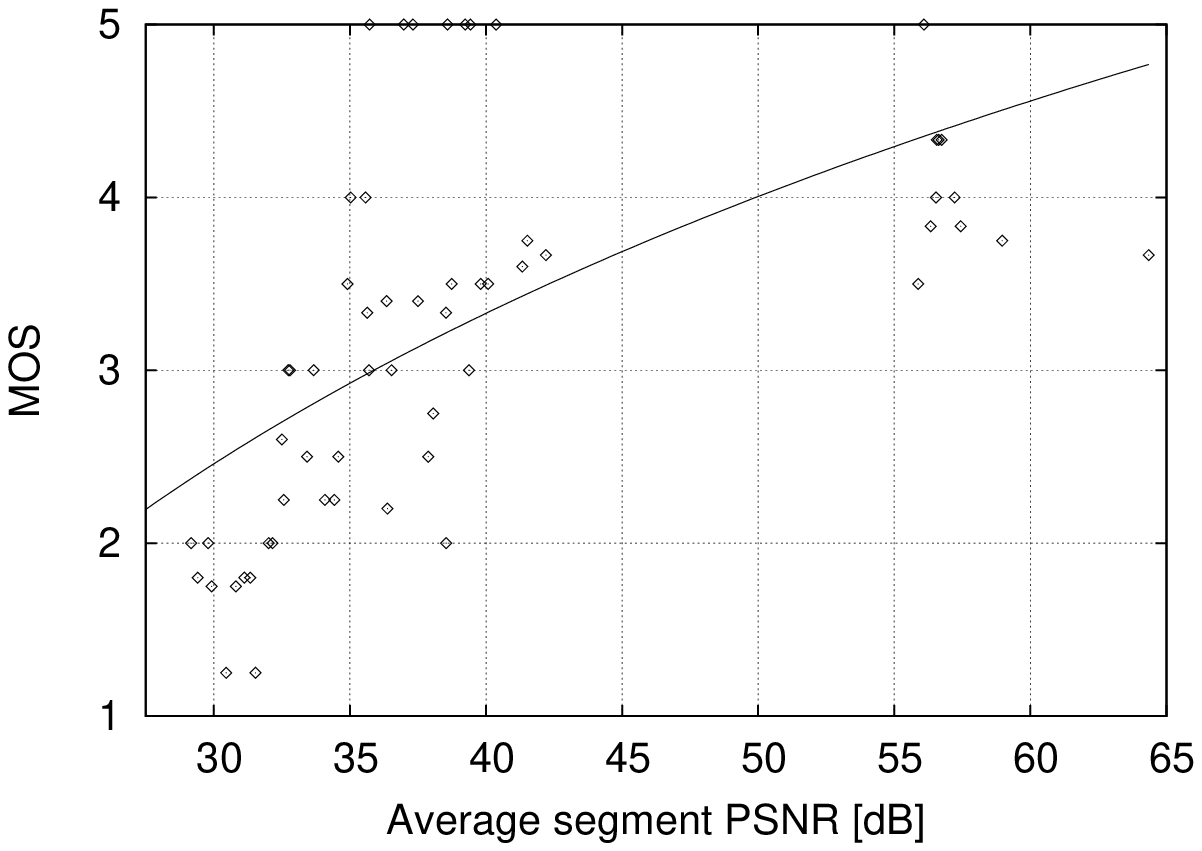}
	}
	\\
	\subfloat[SSIM\label{sfig:mosssim}]{%
		\includegraphics[height=0.3\textheight]{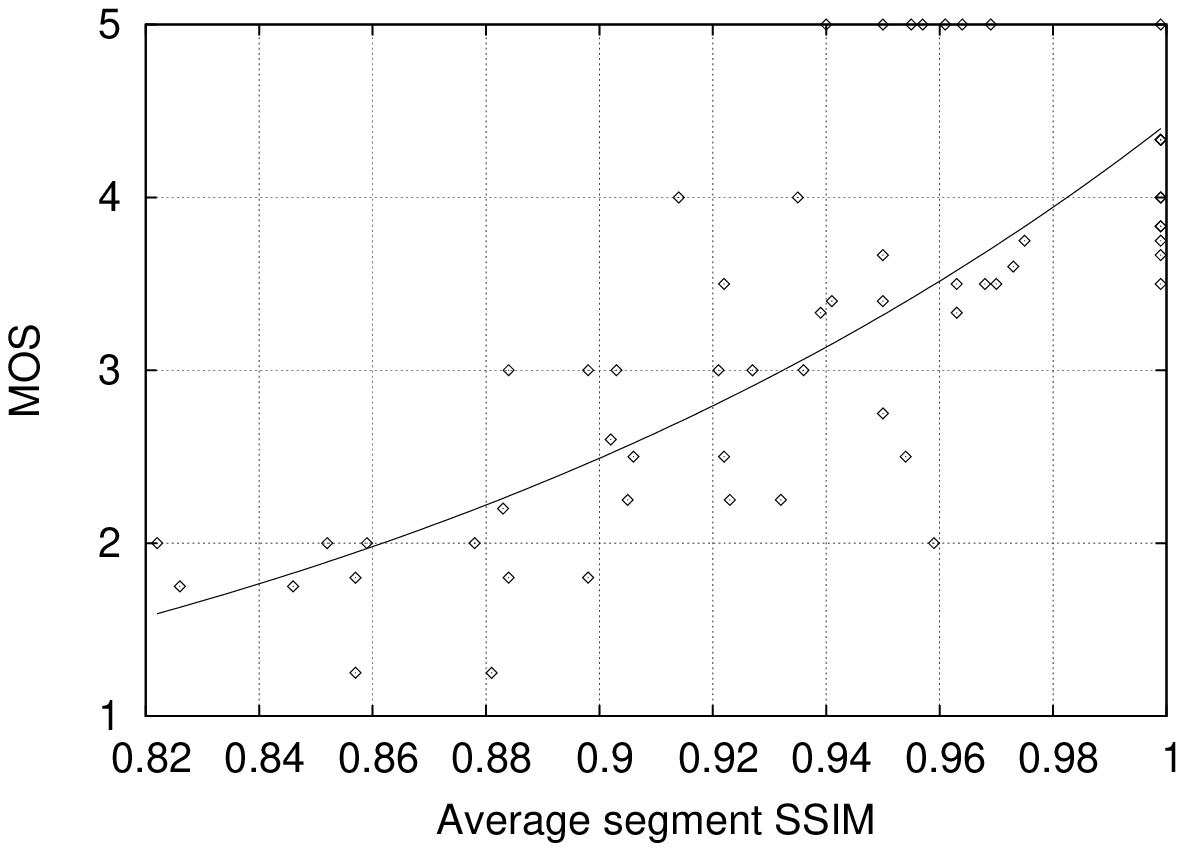}
	}
\\
	\subfloat[VQM\label{sfig:mosvqm}]{%
		\includegraphics[height=0.3\textheight]{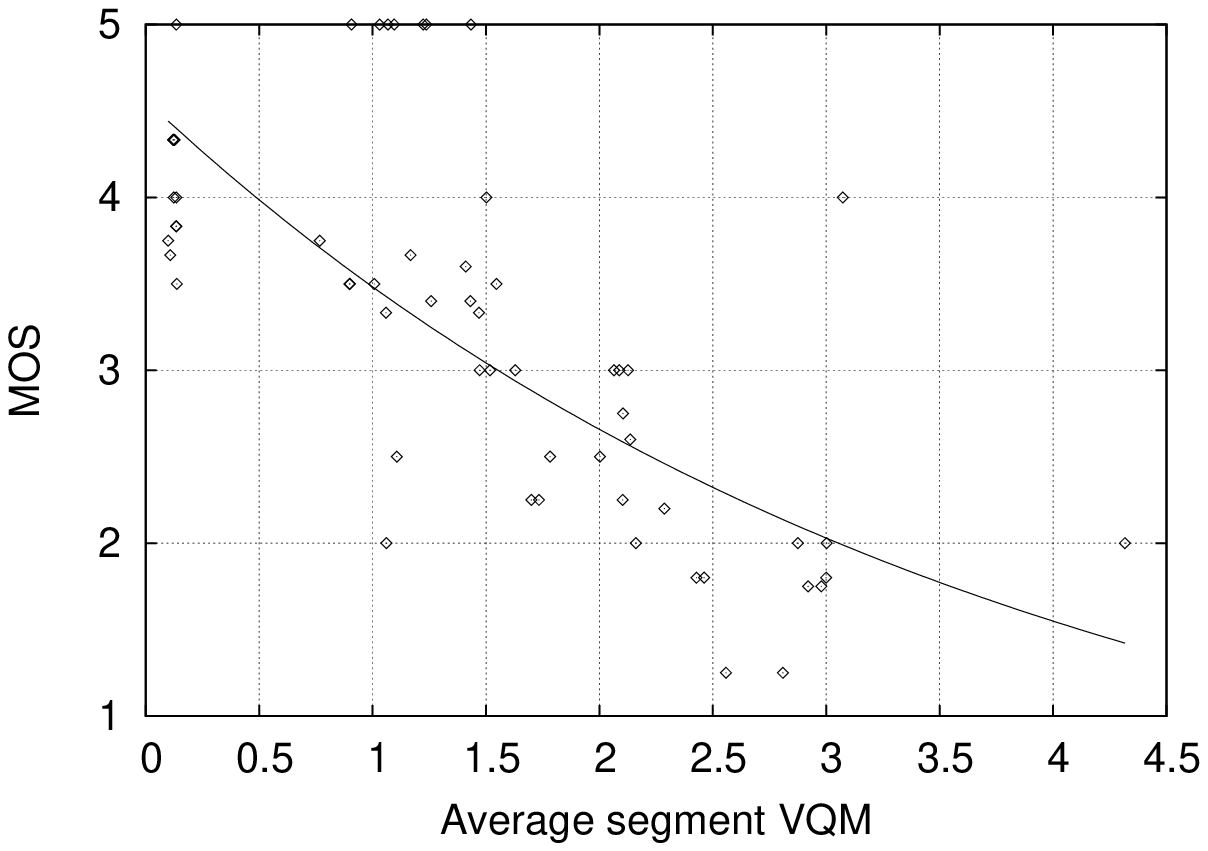}
	}
	\caption{Mean opinion scores (MOS) determined from segment averages of the PSNR, SSIM, and VQM metrics.}
	\label{fig:mos_scores}
\end{figure}

Rather than just relying on our classification of the video sequences by modifying a constant quantization scale, we additionally present the mean opinion score as result of the underlying video segment qualities a participant experienced as part of the study. 
We employ the PSNR, SSIM, and VQM metrics as previously described in Section~\ref{ss:vieo_desc}. We illustrate these relationships in Figure~\ref{fig:mos_scores}.
We initially observe that the different encoding settings result in a widespread range of qualities indicated by the average segment PSNR values. 
	Furthermore, due to our initial high quality setting, these encoded segments stand out a separate cluster towards the high end of PSNR values.
	For the medium range, we observe the significant number of high ratings performed by the participants.
	Evaluating the overall relationship, we derive a fitted logarithmic relationship of 
	\begin{equation*}
		\mathrm{MOS} = 3.028 \cdot ln(\mathrm{PSNR}) - 7.8402,
	\end{equation*}
which, due to the variability of user ratings, resulted in $R^2 = 0.366$.
Next, we evaluate the interplay of the MOS and the average segment SSIM, illustrated in Figure~\ref{sfig:mosssim}.
	As the similarity--based metric is related to the participant opinions, we note that the overall spread is somewhat reduced visibly, which is attributed to the high ratings of participants towards the higher SSIM values, as well as the SSIM values approaching one for the high quality (i.e., low quantization scale) encodings.
	We note that the relationship tends to be inverse to the one observed for the PSNR in Figure~\ref{sfig:mospsnr}.
	Indeed, a fitted curve indicates an exponential relationship described by 

\begin{equation*}
\mathrm{MOS} = 0.0142 \cdot e^{5.7414 \cdot \mathrm{SSIM}},
\end{equation*}
resulting in a higher $R^2 = 0.59993$.
	
Thirdly, we consider the MOS in relationship to the average segment VQM values in Figure~\ref{sfig:mosvqm}.
	We initially observe an almost inverse relationship between the VQM and SSIM, due to one rating similarities, the other one rating negative impacts on quality. 
	We additionally note that a slightly more condensed relationship between viewer ratings and average VQM values can be observed for the medium ranges.
	The fitted declining function can be expressed as
\begin{equation*}
\mathrm{MOS} = 4.5622 \cdot e^{-0.27 \cdot \mathrm{VQM}},
\end{equation*} 
with a resultant $R^2 = 0.51815$.

Overall, the relationships between the participant--reported MOS scores and the objective indicators measured by PSNR, SSIM, and VQM metrics are in line with prior findings between QoS and QoE, such as those reported in \cite{LogLaw2013}, \cite{IQX2010}.
We note, however, that additional studies are required to reduce the currently inherent variability in participant reported MOS values.

\subsection{Comparing Video Quality Ranking and MOS Results}
Comparing the means through an ANalysis Of VAriance (ANOVA) to test for difference of means reveals that they are to be considered related $F(1,354) = 1.89, p > 0.169$. 
For all pairs, independently of the segment, we observe a correlation of $\rho_{v,p}=0.62$, which indicates a possible relationship between the encoded video quality level and user--identified one. 
Next, we consider the relationship within the individual segments to derive a more detailed view on the participant selection given content and segment length differences.
We compare the pair--wise correlation between the set and user--selected video quality values on the 5--point scale and their two--tailed T--Test significance (for a 95 \% confidence level) in Table~\ref{tab:correl}.
\begin{table}[]
\centering
\caption{Overview of correlation and paired sample T--Test two--tailed significance values for segment--based video quality levels and user selections.}
\label{tab:correl}
\begin{tabular}{|c|c|c|c|c|c|c|}\hline
Segm. & Samples & Diff. Avg. & Diff. Std. Dev. & Corr. & T & Sign. \\
$s$ 	& $N$	& $\mu_{v-p}(s)$ & $ \sigma_{v-p}(s)$	& $\rho_{v,p}(s)$ & $t$ & $p$\\\hline
1	&15	 &  -0.067	&	1.486			& 0.562	&    -0.174 & 0.865						\\\hdashline[1pt/1pt] 
2	&15	 &  -0.2	  &	1.568			&	 0.446	&    -0.494	& 0.629					\\\hdashline[1pt/1pt] 
3	&15	 &  -0.267	&	0.884			& 0.809	&    -1.169	& 0.262						\\\hdashline[1pt/1pt] 
4	&15	 &  -0.267	&	1.163			& 0.563	&    -0.888	& 0.389						\\\hdashline[1pt/1pt] 
5	&14	 &  0.071	&		0.997		& 0.725	&    0.268	& 0.793						\\\hdashline[1pt/1pt] 
6	&15	 &  -0.067	&	1.387			& 0.255	&    -0.186	& 0.855						\\\hdashline[1pt/1pt] 
7	&15	 &  -0.267	&	0.961			& 0.798	&    -1.075	& 0.301						\\\hdashline[1pt/1pt] 
8	&15	 &  -0.4	  &	1.121				& 0.603	&    -1.382	& 0.189						\\\hdashline[1pt/1pt] 
9	&15	 &  -0.467	&	1.06				& 0.718	&    -1.705	& 0.11						\\\hdashline[1pt/1pt] 
10&	15 &  0.133		&	1.187				& 0.720	&    0.435	& 0.67						\\\hdashline[1pt/1pt] 
11&	15 &  -0.267	&	1.163				& 0.554	&   -0.888	& 0.389						\\\hdashline[1pt/1pt] 
12&	14 &  -0.286	&	1.383				& 0.368	&   -0.773	& 0.453						\\\hline
\end{tabular}
\end{table}
We initially observe that with exceptions for segments 5 and 10, the difference average is slightly negative, indicating that on average, participants chose higher quality levels than displayed.
The comparatively large standard deviation indicates that users deviate significantly from the actual displayed values in almost every segment, with the exceptions of segments 8 and 9.
These two segments exhibits higher levels of content dynamics as the plot of the movie moves towards its climax.
We note that the correlation is with few exceptions over 0.5, indicating again that user--selected values and randomly displayed video quality levels are potentially related.
We compare these findings by performing paired T--Tests for the individual user selections in each segment and present results in Table~\ref{tab:correl} as well.
The relatively small differences in average, paired with the calculated standard deviations, do not indicate that there is a statistically significant difference between the video categories presented and the ones that were participant--selected, which is corroborated by the $p$--values obtained for the individual segments.
The smallest $p$--value determined is 0.11, which is slightly above typical significance levels.

\section{Selection Performance}
\label{s:selection}
In this section, we interpret the selection of the video quality by participants as a retrieval process and calculate the typical performance measures.
We denote the user--selected quality level $u$ and the randomly displayed encoded video quality level $v$ for each segment $s$ as in the preceding Section~\ref{s:results} and provided in Table~\ref{tab:selections}.
We note that throughout this section, we assume that the the video quality level $v$ is at least a ranking--wise close representation of the grounded truth.
	Commonly, a determination of the ground truth requires extensive human subject ratings to allow for a broad judgment base.
	Here, we assume that the ranking of video qualities as performed based on the different metrics is a reflection of the ground truth. 
	This assumption leans itself onto prior comparisons of full reference metrics, such as \cite{Sheikh06} for still images or \cite{Aggarwal14} for applications.

\subsection{Metrics}
We employ the common notation introduced in, e.g., \cite{Sokolova:2009hy}, by defining the confusion matrix in dependence of a specific video quality level $\nu$ (where $[\cdot]$ denotes the Iverson Bracket) as follows:
\begin{eqnarray}
\mathrm{True Positive} && tp(\nu) = \sum_{s,u} \left[ v(s,u) = \nu \right] \cdot \left[ u(s,u) = \nu \right]\\
\mathrm{False Positive} && fp(\nu) = \sum_{s,u} \left[ v(s,u) \ne \nu \right] \cdot \left[ u(s,u) = \nu \right]\\
\mathrm{False Negative}&&  fn(\nu) = \sum_{s,u} \left[ v(s,u) =\nu \right] \cdot \left[ u(s,u) \ne \nu \right]\\
\mathrm{True Negative}&&  tn(\nu) = \sum_{s,u} \left[ v(s,u) \ne \nu \right] \cdot \left[ u(s,u) \ne \nu \right]
\end{eqnarray}
Omitting the relationship to $\nu$ for clarity, the common performance metrics are defined as:
\begin{eqnarray}
\mathrm{Accuracy} && acc = \frac{tp+tn}{tp+fp+fn+tn}\\
\mathrm{Precision} && prec = \frac{tp}{tp+fp}\\
\mathrm{Recall} && rec = \frac{tp}{tp+fn} \\
\mathrm{F-Score} && F_1 = \frac{2 \cdot prec \cdot rec}{prec + rec}
\end{eqnarray}
We employ these values to determine the performance of the participant selection of a displayed video quality as result of the human quality perception in relationship to the encoded video quality levels for the individual segments.

\subsection{Video Quality Dependence}
\begin{figure}
\centering
\includegraphics[width=0.65\linewidth,angle=270]{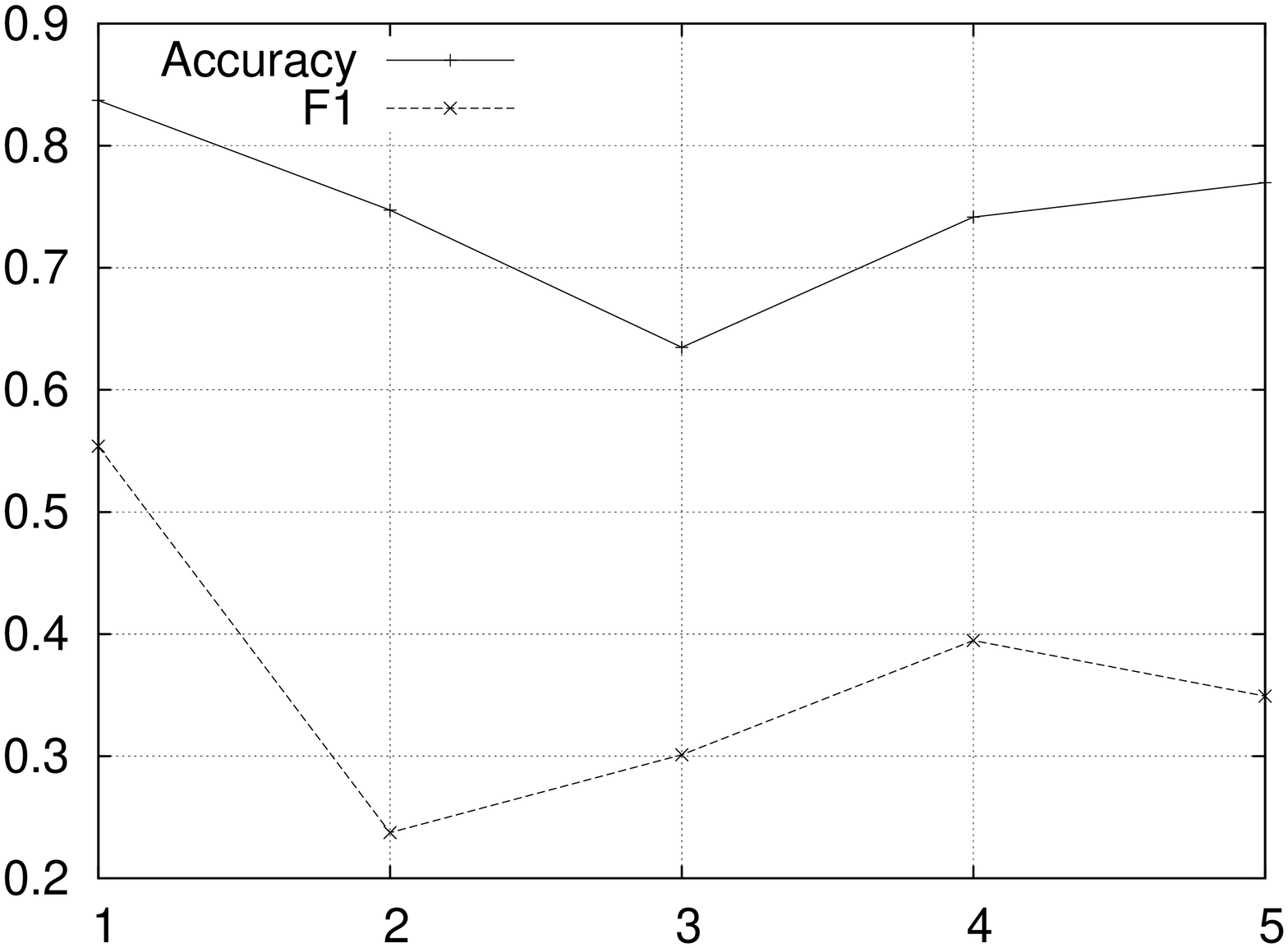}
\caption{Selection performance results in terms of accuracy and $F_1$-score depending on the video quality level $v$. Medium video quality level ranges result in lower participant selection performance.}
\label{fig:all}
\end{figure}

We initially note an overall average accuracy in video quality selection of $acc=75.6$ \% (or error rate of 25.4 \%), indicating that for the majority of segments, participants were able to discern the video quality without training correctly into one of the five quality levels.
The precision and recall values observed are $prec=40.8$ \% and $rec=35.9$ \%, respectively, resulting in an F-score of $F_1 = 0.38$.
This indicates that overall, users were only exhibiting low--medium ability to correctly identify the video quality level.
The error rate can be explained by the nature of the see--through display, which might allow certain types of video quality impairments to go unnoticed.
The dependency of the different values becomes more apparent when evaluating the user performance in dependency of the underlying video quality, as illustrated in Figure~\ref{fig:all}.

We observe that the accuracy and $F_1$ scores both start on a high level, decrease with higher quality levels, followed by an additional increase. 
Only the $F_1$ score exhibits a slight decrease for the highest quality.
The accuracy is the lowest for the medium quality, which can be explained with parts of sequences exhibiting higher levels of complexity, which result in higher levels of compression artifacts even in medium quality settings.
As a result, participants are rating the displayed quality lower than it actually is; opposite considerations apply for a better quality rating.
At the extreme ends, there are either significant quality impairments throughout a segment or only very few, which likely makes it relatively easy to discern these endpoints and, thus, results in higher accuracy.

\subsection{Segment Dependence}
\begin{figure}
	\centering
	\includegraphics[width=0.65\textwidth,angle=270]{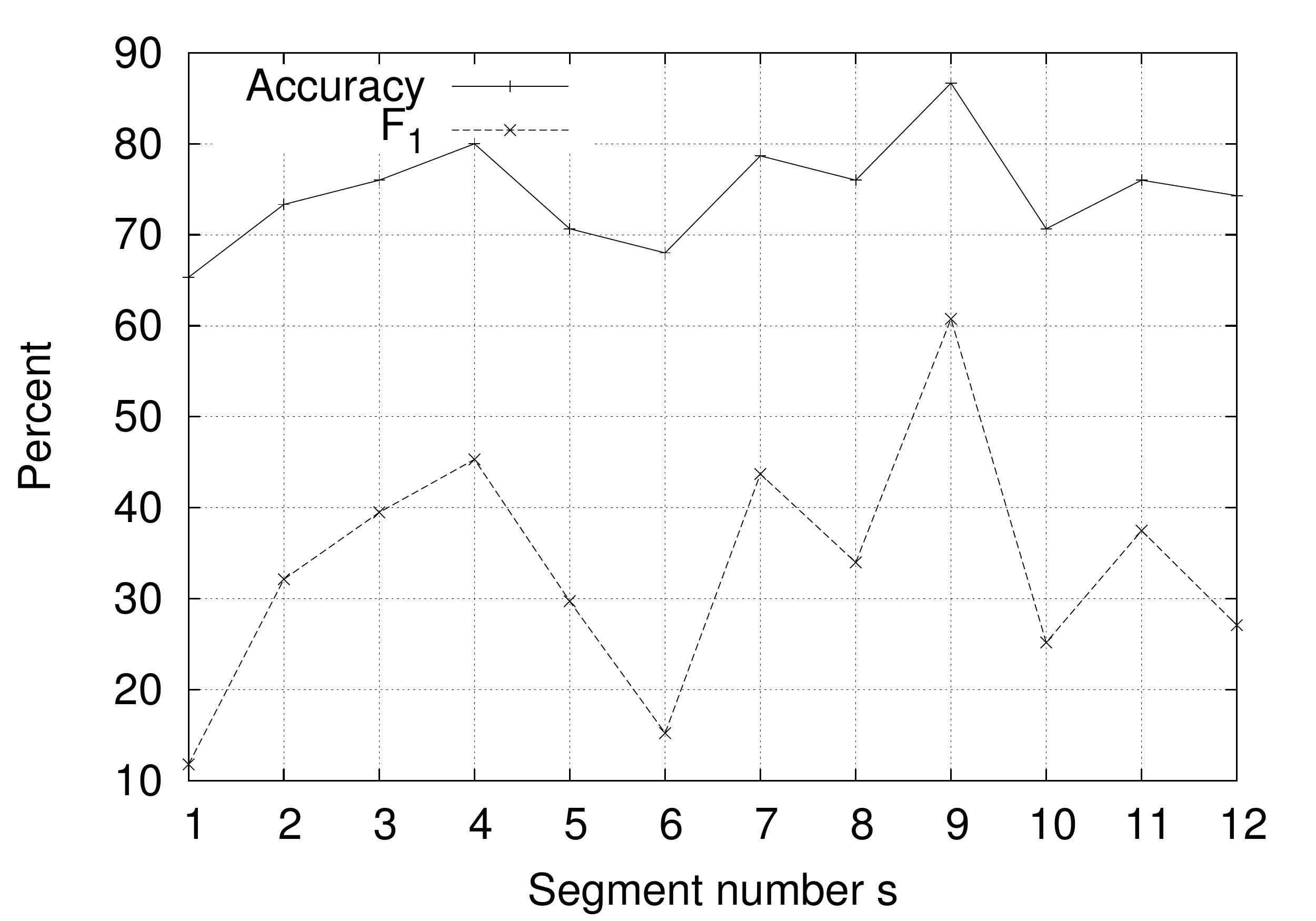}
	\caption{Segment--dependent participant selection performance. Selection performance coincides with overall movie content and storyline dynamics.}
	\label{fig:segments}
\end{figure}

We now shift the view to evaluate the impact of the content present in the different segments on the accuracy and $F_1$ score of the participants' selection of the video quality when compared to the actual ones, illustrated in Figure~\ref{fig:segments}.

We observe an average accuracy that overall remains around or above 70 \%. 
We note an initial rise, followed by a drop to the middle of the complete movie, followed by an increase and a final decrease.
This behavior is followed closely by the $F_1$ score as well, but with larger differences in the rising and falling trends.
Segment 9 exhibits the highest values for both, with an accuracy above 85 \% and an $F_1$ score just above 60 \%.
As a segment with several highly dynamic action--scenes, the imperfections become more obvious, e.g., pixelations or blockiness in explosions.
However, the rise to this point also coincides with the tension of the actual movie (that climaxes in segments 9 and 10), which might be an additionally contributing factor.
Overall, these results indicate that content variation has an impact on accuracy and precision/recall and needs to be considered as in regular display facilities for video encodings.

\section{Conclusion}
\label{s:conc}

The mobile consumption of movie content in augmented reality settings gives rise to the question of how mobile users perceive the display of multimedia content on their devices; here, we presented the the first study addressing this research domain for wearable binocular vision see--through displays using a commercial off-the--shelf consumer device.

For the publicly available \emph{Tears of Steel} short movie, segmented into multiple shorter sequences, we find that users tend to slightly overestimate the video quality, with no statistically significant difference of means (but approaching it for individual segments).
The participant--selected high quality levels tend to correlate with the content of the segments, with higher levels of content dynamics exhibiting larger positive ratings compared to the presented video quality level. 
Though overall, we notice a medium--high accuracy level around 75 \%, the precision and recall values are significantly lower, corroborating the general results.
We reason that a significant portion of the positive viewer bias stems from the nature of the optical see-through device, which likely obscures smaller visual imperfections when compared to a traditional display method.
This is substantiated by participant selections exhibiting higher levels of accuracy, precision, and recall for high and low video quality levels throughout, but lower values in the medium range, where some obfuscation might lead to higher quality ratings.

Future multimedia delivery systems targeting this form of media display can take these findings into account to optimize content modification and delivery mechanisms.
A necessary refinement required for future evaluations of content characteristics, compression, delivery, and adaptation methods is the determination of a detailed testing protocol that allows researchers to perform comparable evaluations. 
With a consensus on such a protocol, future investigations in this domain will become enabled to target fine--grained parameters commonly employed in today's traditional display settings.
Future research avenues can evaluate more interplays of audio quality or ``background'' real--world settings and their influence on the perceived quality.

\section*{Acknowledgments}
We thank Joshua Whaley for his development efforts of the mobile application.
Sponsored in part by an Early Career grant from the Office of Research and Sponsored Programs at Central Michigan University.

\end{document}